\documentclass[aps,prl,superscriptaddress,twocolumn,showpacs]{revtex4-1}

\pdfoutput=1

\usepackage[utf8]{inputenc}
\usepackage[english]{babel}
\usepackage{setspace}  
\usepackage{graphicx}
\usepackage{subfigure}
\usepackage{amssymb}
\usepackage{amsmath}
\usepackage{amscd}
\usepackage{rotating}
\usepackage{color}
\usepackage{epstopdf}
\usepackage{enumerate}

\usepackage{hyperref}
\hypersetup{
        colorlinks=true,
}

\usepackage{amsthm}

\begin{document}

\newcommand{\BIGOP}[1]{\mathop{\mathchoice%
{\raise-0.22em\hbox{\huge $#1$}}%
{\raise-0.05em\hbox{\Large $#1$}}{\hbox{\large $#1$}}{#1}}}
\newcommand{\bigtimes}{\BIGOP{\times}}

\renewcommand{\v}[1]{\ensuremath{\mathbf{#1}}} % for vectors
\newcommand{\gv}[1]{\ensuremath{\mbox{\boldmath$ #1 $}}} % for vectors of Greek letters
\newcommand{\md}[1]{\mathrm{d}#1\,} % for differential
\renewcommand{\d}[2]{\frac{d #1}{d #2}} % for derivatives
\newcommand{\dd}[2]{\frac{d^2 #1}{d #2^2}} % for double derivatives
\newcommand{\pd}[2]{\frac{\partial #1}{\partial #2}} % for partial derivatives
\newcommand{\pdd}[2]{\frac{\partial^2 #1}{\partial #2^2}} % for double partial derivatives
\newcommand{\pdc}[3]{\left( \frac{\partial #1}{\partial #2} \right)_{#3}} % for thermodynamic partial derivatives
\newcommand{\ket}[1]{\left| #1 \right>} % for Dirac bras
\newcommand{\bra}[1]{\left< #1 \right|} % for Dirac kets
\newcommand{\braket}[2]{\left< #1 \vphantom{#2} \right| \left. #2 \vphantom{#1} \right>} % for Dirac brackets
\newcommand{\matrixel}[3]{\left< #1 \vphantom{#2#3} \right| #2 \left| #3 \vphantom{#1#2} \right>} % for Dirac matrix elements

\newcommand{\comment}[1]{\emph{\color{red} #1}}

\title{Multigrid Algorithms for Tensor Network States}

\author{M. Dolfi}
\affiliation{Theoretische Physik, ETH Zurich, 8093 Zurich, Switzerland}

\author{B. Bauer}
\affiliation{Station Q, Microsoft Research, Santa Barbara, CA 93106-6105, USA}

\author{M. Troyer}
\affiliation{Theoretische Physik, ETH Zurich, 8093 Zurich, Switzerland}

\author{Z. Ristivojevic}
\affiliation{Laboratoire de Physique Th\'eorique-CNRS, Ecole Normale Sup\`erieure, 24 rue Lhomond, 75005 Paris, France}

\date{\today}

\begin{abstract}
The widely used density matrix renormalization group (DRMG) method often fails to converge in systems with multiple length scales, such as lattice discretizations of continuum models and dilute or weakly doped lattice models. The local optimization employed by DMRG to optimize the wave function is ineffective in updating large-scale features. Here we present a multigrid algorithm that solves these convergence problems by optimizing the wave function at different spatial resolutions. We demonstrate its effectiveness by simulating bosons in continuous space, and study non-adiabaticity when ramping up the amplitude of an optical lattice. The algorithm can be generalized to tensor network methods, and be combined with the contractor renormalization group (CORE) method to study dilute and weakly doped lattice models.
\end{abstract}

% 02.70.-c	Computational techniques; simulations
% 03.65.Ud	Entanglement and quantum nonlocality (e.g. EPR paradox, Bell's inequalities, GHZ states, etc.) (for entanglement production and manipulation, see 03.67.Bg; for entanglement measures, witnesses etc., see 03.67.Mn; for entanglement in Bose-Einstein condensates, see 03.75.Gg)

% 37.10.Jk	Atoms in optical lattices

% 67.85.De	Dynamic properties of condensates; excitations, and superfluid flow
% 67.85.Hj	Bose-Einstein condensates in optical potentials

% 71.27.+a	Strongly correlated electron systems; heavy fermions

\maketitle

% INTRO DMRG
The optimization of variational wave functions is generally a very difficult problem. In the specific case of matrix-product states (MPS)~\cite{ostlund1995} the density-matrix renormalization group algorithm~\cite{white1992,white1992-1,schollwoeck2005,schollwock2011} often reliably and efficiently optimizes these wave functions to find a good approximation of the ground state. While most efficient in one dimension, it can  be  applied to medium-sized two-dimensional systems~\cite{stoudenmire2012}, and has been generalized to calculate time-dependent \cite{daley2004,vidal2003,white2004} and finite temperature properties \cite{verstraete2004-mpdo,feiguin2005,white2009}. 

% PROBLEMS / EXAMPLES
In systems with multiple length scales, however, the DMRG algorithm often fails to converge, as the local optimizations that are at the core of DMRG are ineffective in optimizing large-scale features of the wave function. Especially in dilute systems where the inter-particle distance is large compared to the lattice spacing the convergence of the density profile can be very slow. Systems with multiple length scales suffering from this problem arise from lattice discretizations of continuum models~\footnote{While a recently developed approach allows the simulation of continuum systems without introducing a lattice~\cite{verstraete2010}, the standard DMRG method on a fine mesh is currently more robust and accurate \cite{dolfi2011}.}, or in weakly doped lattice models where the hole density exhibits the same convergence problems. The first situation was recently discussed in Ref.~\cite{stoudenmire2011}.

\begin{figure}[t]
\centering
{\small
\includegraphics{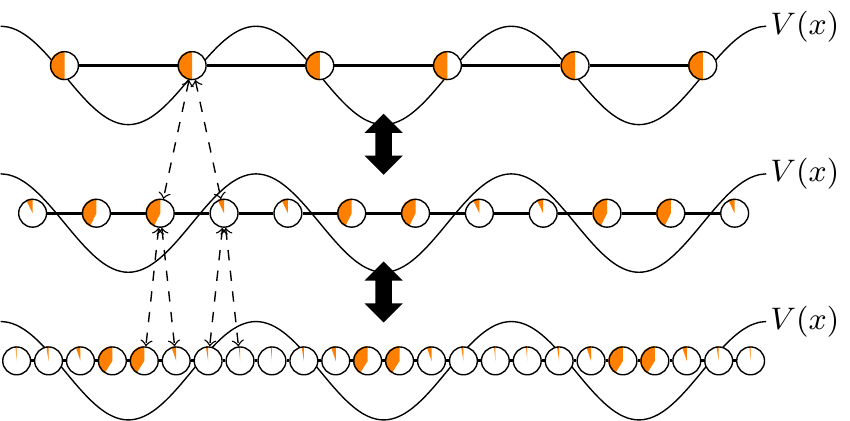}
}
\caption{Multigrid DMRG illustrated for bosons in an optical lattice $V(x)$. The DMRG algorithm converges fast for the rather dense system at the coarsest grid (shown on the top), and this solution is then used to iteratively initialize DMRG calculations on finer grids, which substantially speeds up convergence. The filling of the circles illustrates the probability for a particle to be at that site.}
\label{fig:lattice-fine-graining}
\end{figure}

% OTHER FIELDS AND multigrid (in general)
Similar convergence problems are also known in other fields, {\it e.g.}~when solving partial differential equations~\cite{stuben2001}, lattice field theories~\cite{goodman1986} or electronic structures~\cite{heiskanen2001}, and have there been overcome by {\em multigrid} approaches.  Multigrid methods use a hierarchy of discretizations, as sketched in Fig.~\ref{fig:lattice-fine-graining}. Starting from the target problem on the finest grid (or a lattice model), the system is mapped to hierarchy of coarser grids. An approximate solution of the smallest problem on the coarsest grid is then used to initialize  optimizations of the problem on the next finer grid and this process is iterated down to the finest grid. This method can substantially speed up a calculation since the large scale features converge quickly on the coarsest grid, and the following calculations on finer grids only need to optimize local features at the scale of the respective grid spacing.

% WHAT WE DO (new method+applications)
In this Letter we develop a multigrid DMRG (MG-DMRG) algorithm to solve the above-mentioned convergence problems in DMRG calculations.  As a  first application and demonstration of the effectiveness of the algorithm we present results for bosons in continuous space where MG-DMRG enables the study of non-adiabaticities when slowly ramping up the amplitude of an optical lattice.

% ALGORITHM
We start the description of the MG-DMRG algorithm by reviewing MPS wave functions on a chain of $L$ sites:
\begin{equation}
\label{eq:mps}
\ket{\psi} = \sum_{\gv \sigma} A_1^{\sigma_1} A_2^{\sigma_2} \cdots A_L^{\sigma_L} \ket{\gv \sigma}
\end{equation}
used in DMRG. They are characterized by a polynomial number $\propto LM^2$ of variational parameters, the $M\times M$ matrices $A_i^{\sigma_i}$. In one dimension a good approximation for low-energy states can be obtained by MPS wave functions with a fixed or at most polynomially growing $M$~\cite{verstraete2006-2,schuch2008-1}.

\begin{figure}[tb]
\vspace{-1em}
\includegraphics[width=\linewidth]{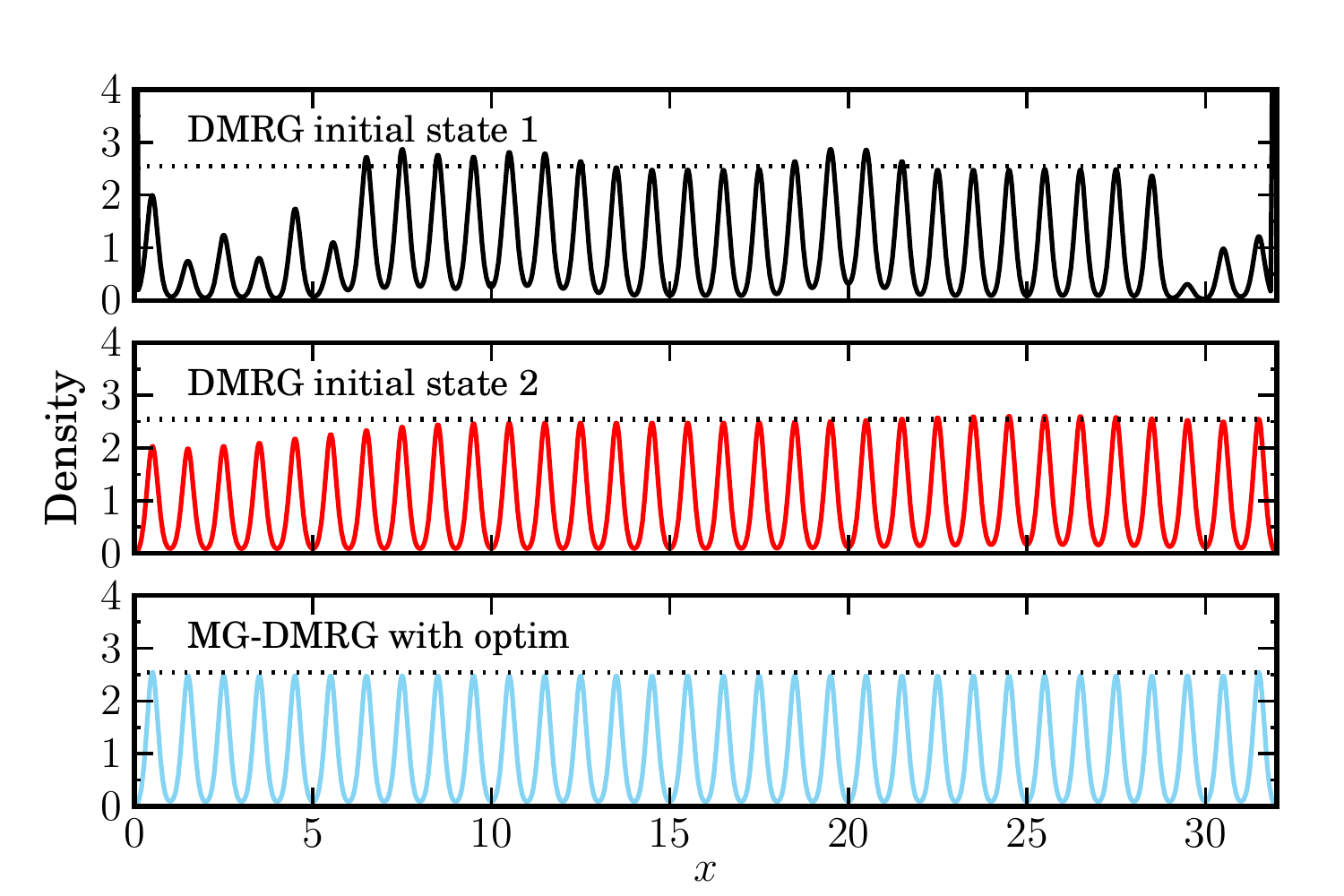}
\caption{Density profiles of a continuum system of bosons in an optical lattice consisting of $L=32$ unit cells. The top and middle panels shows the non-converged results obtained for $N=32$ grid points per unit cell after $12$ sweeps of the DMRG algorithm with two different initial states: initial state 1 is a random state, initial state 2 is obtained from an infinite-size growing procedure as implemented in the ALPS DMRG code~\cite{bauer2011-alps}. The bottom panel shows the multigrid result.
}
\label{fig:density}
\end{figure}

While optimizing MPS wave functions to obtain a variational estimate for the ground state is a hard non-linear problem, the DMRG algorithm is very effective in many cases. It iteratively optimizes one or two of the matrices  $A_i^{\sigma_i}$ while keeping all other matrices fixed, and sweeps back and forth along a (quasi) one-dimensional system until convergence is achieved. For a recent review and implementation see Ref. \cite{schollwock2011}. It can, however, get trapped in local minima of this non-linear optimization problem, or become very slow especially for the dilute systems considered here. As an example see the badly converged density profiles obtained by standard DMRG approaches in Fig. \ref{fig:density}.

In our implementation of the MG-DMRG algorithm, we start by constructing the target lattice model and a hierarchy of models on coarser grids. Starting from the coarsest level we optimize the wave function and interpolate it to the next finer level, repeating this procedure until we reach the target system. Many generalizations are possible, for example iterating the procedure by going back to coarser levels, or starting from the finest instead of coarsest level. 
%The MG-DMRG algorithm performs the following steps. First the target lattice model and a hierarchy of models on the coarser grids are constructed.\comment{I would prefer to rephrase this to contain the case where the coarser models can't immediately be constructed, for example if the isometry is obtained by some reduction of the state} Starting from the coarsest level one optimizes the wave function and interpolates it to the next finer level, repeating this procedure until one reaches the target system. The wave function can then again be mapped back to the coarser levels and the optimization repeated to achieve an even better solution.

\begin{figure}[b]
\centering
{\small
\includegraphics{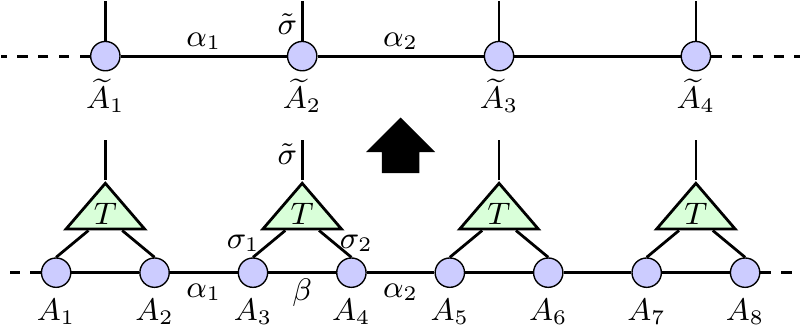}
}
\caption{The {\em restriction} of an MPS state reducing it from 8 to 4 sites. The contraction along the indices $\beta$, $\sigma_1$ and $\sigma_2$ creates a new tensor $\widetilde A^{\widetilde \sigma}_{\alpha_1\, \alpha_2}$ with a new index $\tilde \sigma$, but keeping the old connections to the neighboring sites ($\alpha_1$ and $\alpha_2$).}
\label{fig:coarse-graining}
\end{figure}

The {\em restriction} operation maps a system to a coarser grid, merging $n$ (typically $n=2$) sites into one. The model, given by the Hamiltonian $H$ and defined in the local basis $\lbrace \sigma \rbrace$, is mapped to a restricted model $\widetilde H$ in a truncated local basis $\lbrace \widetilde \sigma \rbrace$ for the $n$ merged sites. The truncation, denoted by an isometry $T^{\widetilde \sigma}_{\sigma_1 \dots \sigma_n}$, is straightforward for continuum models and an approach for lattice models will be discussed below. Any error due to the truncation will be corrected when returning to finer scales, as long as we stay in the same phase. As illustrated in Fig.~\ref{fig:coarse-graining} the restriction transforms a matrix product state $A_1$, \dots, $A_n$ into 
\begin{equation}
\label{eq:coarse-graining}
\widetilde A^{\widetilde \sigma}_{\alpha_1\, \alpha_2} = A^{\sigma_1}_{\alpha_1\, \beta_1}\; A^{\sigma_2}_{\beta_1\, \beta_2} \cdots \; A^{\sigma_n}_{\beta_{n-1}\, \alpha_2} \; T^{\widetilde \sigma}_{\sigma_1 \dots \sigma_n}.
\end{equation}

\begin{figure}[b]
\centering
{\small
\includegraphics{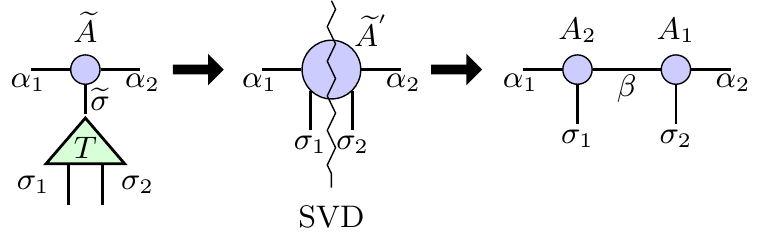}
}
\caption{{\em Prolongation} operation doubling the number of sites: first we transform the basis $\widetilde \sigma$ of the initial matrix $\widetilde A$ into two new local bases $\sigma_1$ and $\sigma_2$. With a singular value decomposition we then split the rank-$4$ tensor $\widetilde{A}^{'}$ into two matrices $A_1$ and $A_2$.}
\label{fig:fine-graining}
\end{figure}

{\em Prolongation} is the inverse of restriction and maps a solution from a coarse grid to a finer one. The isometry is inverted and  $T^{-1}$ replaces one index by  $k$ new indices. From this tensor we can recover a (not unique) MPS representation by repeatedly applying singular value decompositions and  splitting it into matrices $A_1$, \dots, $A_n$ (see Fig.~\ref{fig:fine-graining}). It has turned out to be useful to perform a standard DMRG update on the newly obtained matrices immediately after prolongation while keeping the rest of the system on the coarse-grained lattice~\footnote{This requires a representation of the Hamiltonian on a non-uniform grid.}.

% APPLICATIONS
%  - continuum models
As a first application we apply MG-DMRG to bosons in a one-dimensional continuum system with an external optical lattice potential, $V(x) = V_0 \cos^2(k x)$, with $k=\pi/a$ and $a$ the size of a unit cell. The continuous-space Hamiltonian describing spinless bosons interacting through a $\delta$-potential in a system of $L$ unit cells and length $L a$ is
\begin{align}
\label{eq:cont-hamil}
\hat H =& \int_0^{La}\md{x} \hat\psi^\dagger(x) \left[ -\frac{\hbar^2}{2m}\dd{}{x} + V(x) \right] \hat\psi(x) \notag\\
&+ \frac{g}{2}\int_0^{La}\md{x} \hat\psi^\dagger(x) \hat\psi^\dagger(x) \hat\psi(x) \hat\psi(x),
\end{align}
where a boson is created at position $x$ with the field operator $\hat\psi^\dagger(x)$, satisfying the usual commutation relations. We express energies in units of the recoil energy $E_r = \frac{\hbar^2 k^2}{2m}$. The interaction $g$ is conveniently parametrized by the dimensionless coupling $\gamma = mg / \hbar^2n$, where $n$ is the density. 

In deep optical lattices the low-energy sector of the model can be mapped to an effective single band Hubbard model with one site per unit cell. We are, however, interested also in weak optical lattices and thus discretize the continuum model on a grid with $N$ points per unit cell and  spacing $\Delta x = a / N$. To discretize the Hamiltonian~\eqref{eq:cont-hamil} on this lattice we replace the Laplacian by a second order finite difference approximation and replace field operators by lattice  annihilation and creation operators $\hat\psi^\dagger(x=(i+1/2)\Delta x) = \frac{1}{\sqrt{\Delta x}}\hat c^\dagger_{i}$. We end up with a Hubbard-like model in a spatially varying potential:
\begin{align}
\label{eq:hubbard-hamil}
\hat H(\Delta x) =& -t(\Delta x) \sum_{i} \left[\hat c^\dagger_{i} \hat c_{i+1} + \text{h.c.} \right] + \sum_{i} \mu_{i}(\Delta x) \hat n_{i} \notag\\
&+ \frac{U(\Delta x)}{2} \sum_i \hat n_{i} (\hat n_{i} - 1)
\end{align}
with $t(\Delta x) = (\hbar^2 / 2m) / \Delta x^2$, $U(\Delta x) = g / \Delta x$, and $\mu_i(\Delta x) = V((i+1/2)\Delta x) + 2 (\hbar^2 / 2m) / \Delta x^2$. A similar lattice model can be formulated for fermions.

With this definition of the Hamiltonian for arbitrary  $\Delta x$ the implementation of MG-DRMG is straightforward. The matrix elements of the isometry $T$ are
%$T^{\widetilde \sigma}_{\sigma_1 \dots \sigma_n} = \delta^{\widetilde \sigma}_{\sigma_1 + \dots + \sigma_n}$ (up to normalization), 
\begin{equation}
T^{\widetilde \sigma}_{\sigma_1 \dots \sigma_n} = \frac{ \delta(\widetilde{\sigma}, \sigma_1+\dots+\sigma_n) }{\sqrt{ \sum_{\sigma_1', \dots} \delta(\widetilde{\sigma},\sigma_1' + \dots + \sigma_n') }},
\end{equation}
where $\sigma_i$, $\sigma_i'$ and $\widetilde \sigma$ are particle-number eigenstates and we truncate the maximum occupation of a site at $N_\text{max}$, i.e. $\sigma_i,\widetilde{\sigma} \in \lbrace 0, \dots, N_\text{max} \rbrace$. Due to particle-number conservation, $T$ is a $N_\text{max} \times N_\text{max}^n$ block-diagonal matrix. Note that we start from a coarse-grained lattice and perform only prolongations.

As a benchmark for the multigrid algorithm we consider an optical lattice with $V_0/E_r = 6$ and $1/\gamma=0.1$ corresponding to the insulating phase~\cite{buchler2003} at unit filling. Our MG-DMRG simulation were performed with up to $N = 128$ lattice sites per unit cell ($\Delta x = 0.0078125$) with $N_\text{max}=2$ keeping $M=200$ states and using $12$ sweeps of single-site updates at each level. We also perform standard DMRG simulations starting from either a random initial state, a state obtained from an infinite-size growing procedure, or a few steps of imaginary time evolution (not shown). The infinite-size growing procedure is commonly used to obtain good initial states for one-dimensional systems and has been proven to be very efficient in most cases. We use the implementation of the ALPS DMRG code~\cite{bauer2011-alps}, which performs the growing procedure on a state with very small bond dimension and increases the bond dimension linearly with the number of sweeps thereafter.

Fig.~\ref{fig:density} shows the density profile obtained with the three approaches. Clearly, the standard DMRG approaches are trapped in configurations with a globally non-homogeneous density distribution and further sweeps are not effective in redistributing the particles. The multigrid method, on the other hand, achieves a symmetric distribution, since it performs the optimization of the large-scale features on the initial coarse mesh with just $N=2$ sites per unit cell, where convergence is very fast. Subsequent calculations on finer lattices are initialized with the prolongated solution of the coarser lattice. This is already close to the ground state, and only the local fine-scale features of the wave function need to be optimized.

\begin{figure}[t]
\includegraphics[width=\linewidth]{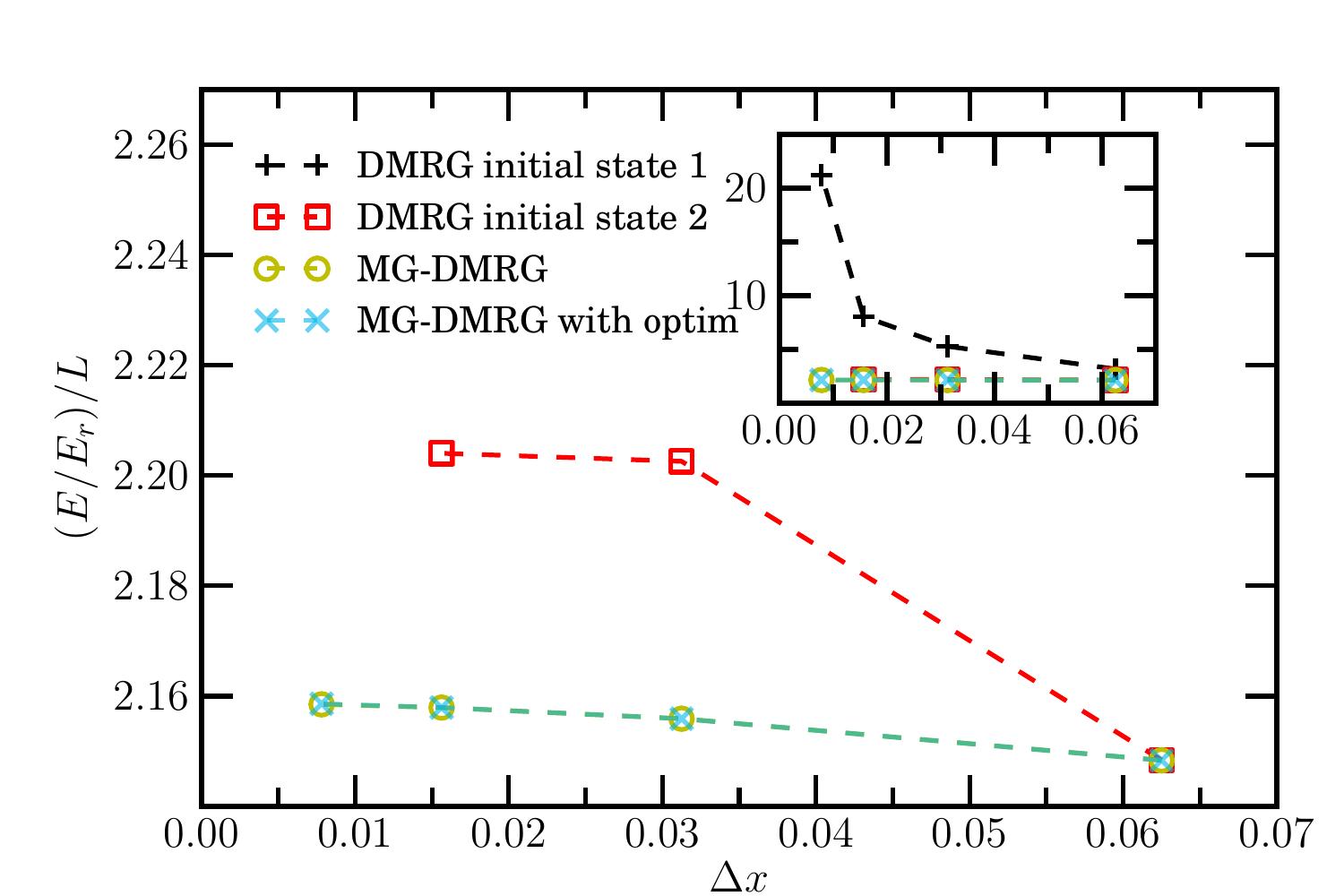}
\caption{Comparison of the energies in a bosonic optical lattice ($V_0/E_r = 6$, $1/\gamma = 0.1$) with $L=32$ unit cells discretized with increasing discretization $N=16,32,64,128$ obtained with different strategies: DMRG with initial state 1 optimizing an initial random state, DMRG with initial state 2 initializing the system with an infinite size procedure and linearly increasing the number of states (results obtained with the ALPS DMRG code~\cite{bauer2011-alps}), MG-DMRG, and MG-DMRG combined with local optimization in the prolongations.}
\label{fig:energy}
\end{figure}

The better convergence of the MG-DMRG is also reflected in the energies shown in Fig.~\ref{fig:energy}. While for modest discretizations, standard approaches yield energies comparable to MG-DMRG, they encounter severe convergence problems for smaller values of $\Delta x$ where the difference between multiple scales of the dilute system become more and more important. The most reliable method is MG-DMRG combined with optimization in the prolongation.

%As expected from the unconverged density profiles, the energies obtained in DMRG are significantly higher than those obtained with MG-DMRG in many cases. Despite a good agreement for a modest discretizations, conventional methods encounter convergence problems for larger more dilute systems, where the difference between multiple scales of the dilute system become more and more important. The best method is MG-DMRG combined with optimization in the prolongation.

\begin{figure}[b]
\vspace*{-1em}
\centering
\includegraphics[width=\linewidth]{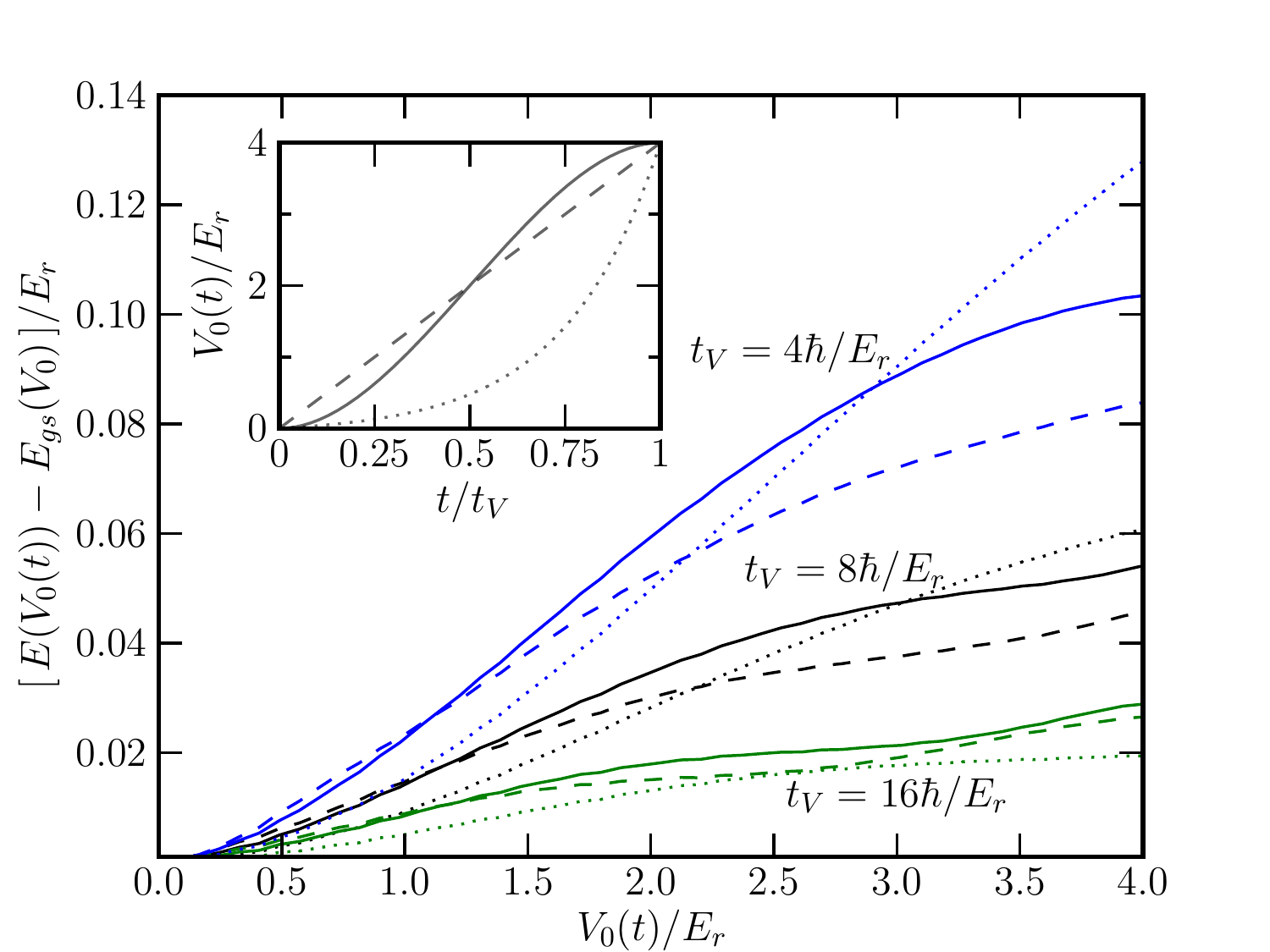}
\caption{Heating due to finite ramping speeds in a system of length $L=16$ for $1/\gamma = 0.08$ with $M=400$. Different lines show the energy difference to the ground state while ramping up to $V_0/E_r=0\rightarrow 4 $. (inset) Ramping functions used in the time evolution: linear, $V_0(t)/E_r = 4 t / t_V$ (dashed); exponential, $V_0(t)/E_r = 4 [\exp(4t/t_V)-1] / [\exp(4)-1]$ (dotted); and $s$-like, $V_0(t)/E_r =4 \left[ 3(t / t_V)^2 - 2(t/t_V)^3 \right]$ (solid).}
\label{fig:optical_ramp}
\end{figure}

MG-DMRG opens new interesting applications for DMRG that have not been accessible before. As an example we combine MG-DMRG with time evolution~\cite{daley2004,vidal2003,white2004}, to study heating caused by non-adiabaticity when ramping up the amplitude of an optical lattice. We start from the ground state of a homogeneous system of length $L=16$ and $N=16$ grid points per unit cell calculated by MG-DMRG. We evolve it in time using a fourth-order Trotter decomposition with $\Delta t = 0.01 \hbar/E_r$. Non-adiabaticities due to ramping at a finite speed cause heating and we plot the energy difference to the ground state in Fig.~\ref{fig:optical_ramp} for three different ramp profiles and several total ramping times. For the calculation of the ground state energies in weak optical lattices MG-DMRG was used. We observe that as the ramp speed decreases, differences in ramp shape are less important than the total ramping time, indicating that the exact shape of the ramp profiles play a minor role in experiments and experimentalists should focus on determining optimal ramping times.

\begin{figure}[t]
\centering
{\small
\includegraphics{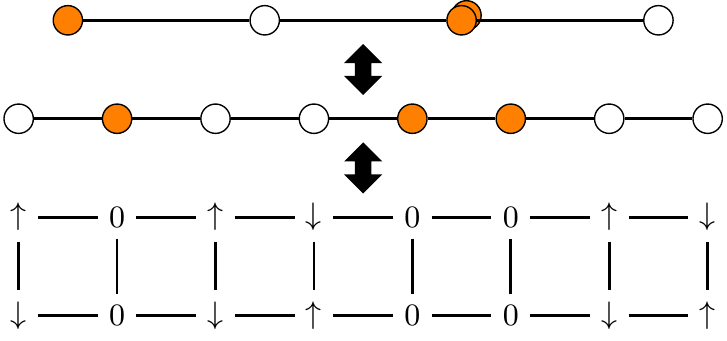}
}
\caption{Coarse-graining a ladder of spin-$1/2$ fermions into a system of hard-core bosons. Spin singlets are mapped to empty sites, and hole-pairs to hard-core bosons in the first step. Further restriction steps of the bosonic model are similar to dilute bosonic models discussed above.}
\label{fig:ladder-coarse-graining}
\end{figure}

% CORE
In DMRG simulations of weakly doped  $t$-$J$ or Hubbard ladder models \cite{Dagotto92,Troyer96,Noack97} the hole density shows similar convergence problems as seen above for dilute particle systems. In particular it has been observed that for six holes in more than $2\times 64$ sites the standard DMRG algorithm fails to distribute the three bound hole pairs evenly over the ladder  \cite{Siller01}, and MG-DMRG can be of use here. The restriction step of mapping the model to a coarser lattice is, however, not as straightforward as in continuum models. We propose to use the contractor renormalization (CORE) method~\cite{core94} to find a good approximation of the model in the reduced Hilbert space of the coarser models, and to iterate this procedure in further restriction steps. For the specific case of doped ladder models, the first step maps 2-site rungs or 4-site plaquettes to a hardcore boson model for the hole pairs, or an extended plaquette model containing hole pairs, magnons, and holes \cite{Altman02}. Further restriction steps map to simpler bosonic models for the hole pairs, as illustrated in Fig. \ref{fig:ladder-coarse-graining}. After prolongation back to the full lattice model the ground state wave function can be further improved by repeating the multigrid scheme can be performed. Now one can use  knowledge of the approximate ground state to perform the restrictions of the basis, instead of using CORE. Details of this method and results of this approach will be published elsewhere.

% COMPARISON WITH TTN & MERA
We point out that MG-DMRG is a fundamentally different approach from the one taken by tree-tensor networks~\cite{shi2006} or the multi-scale entanglement renormalization ansatz (MERA)~\cite{vidal2007-1,vidal2008}. In those approaches, a new class of wave functions is proposed that describes the system at several levels of coarse-graining and all levels are optimized simultaneously -- which can still suffer from  convergence problems at fine scales.  Instead, our approach relies on standard matrix-product states which can be optimized and evaluated much more easily and much faster, but uses a hierarchical coarse graining to achieve a faster and more reliable optimization than standard DMRG.

% OTHER TENSOR NETWORK STATES
Our algorithm can be easily combined with other optimization schemes for DMRG, such as using iTEBD~\cite{vidal2003,vidal2004} to directly simulate  the thermodynamic limit.  One can also easily generalize the restriction~\eqref{eq:coarse-graining} and prolongation to tensors of higher rank, in order to apply the multigrid scheme to other tensor network states, e.g.~MERA~\cite{vidal2007-1,vidal2008}, projected entangled pair states (PEPS)~\cite{verstraete2004} and infinite PEPS (iPEPS)~\cite{jordan2008}.

% Acknowledgment
This project was supported by a grant of ETH Zurich. The calculations were performed with the MAQUIS-DMRG code, developed with support of the Swiss High Performance and High Productivity Computing  (HP2C) initiative, and based on the ALPS libraries \cite{bauer2011-alps}. ZR acknowledges the ANR Grant No.~09-BLAN-0097-01/2.

\bibliography{bibliography}{}
\bibliographystyle{apsrev4-1}

\end{document}